\documentclass{article}

\def\abstract#1{\vskip 7mm 
        \begin{center}{\large Abstract}\par \smallskip
                \begin{minipage}[c]{12cm}
                        \small #1
                \end{minipage}
        \end{center}
}
\def\title#1{\begin{center}{\Large\bf #1}\end{center}}
\def\author#1{\vskip 5mm \begin{center}{#1}\end{center}}
\def\address#1{\begin{center}{\it #1}\end{center}}
\makeatletter
\@ifundefined{lesssim}{}{}
\@ifundefined{gtrsim}{}{}
\def\vereq#1#2{\lower3pt\vbox{\baselineskip1.5pt \lineskip1.5pt
\ialign{$\m@th#1\hfill##\hfil$\crcr#2\crcr\sim\crcr}}}
\makeatother

\begin{document}
\title{%
Fermion Casimir energy for a de Sitter brane in AdS$_5$}
\author{%
Ian G. Moss \footnote{E-mail: ian.moss@ncl.ac.uk}, 
Wenceslao Santiago-Germ\'{a}n \footnote{E-mail: g.w.santiago-german@ncl.ac.uk}
}
\address{%
  \centerline{\em Department of Mathematics, University of  
Newcastle Upon Tyne,} 
\centerline{\em  Newcastle Upon Tyne, NE1 7RU, UK}
}
\author{%
Wade Naylor \footnote{ Current e-mail address: wade@yukawa.kyoto-u.ac.jp}, 
Misao Sasaki \footnote{Current e-mail address: misao@yukawa.kyoto-u.ac.jp}
}
\address{%
\centerline{\em Department of Earth and Space Science, Graduate 
School of Science, Osaka University,} 
\centerline{\em  Toyonaka 560-0043, Japan} 
}
\abstract{
Based on some recent work of the authors, we focus on the 
relationship between the Casimir energy of a Majorana spinor field for a Euclidean 
Einstein universe $S^4\times R$ and for a Euclidean de Sitter brane ($S^4$) 
embedded in AdS$_5$. 
This is for a conformally coupled massless field. 
Interestingly, the one brane effective potential is zero and the results are 
equivalent, as for the scalar case, when evaluated on the conformally related cylinder. 
However, using the actual metric this equivalence no longer holds because a non-trivial 
contribution from the path integral measure (known as the cocycle function) is non-zero.
}
\rightline{\hbox{OU-TAP-196}}
\section{Introduction}

This short talk is based on a small part of some recent (and continuing) work 
\cite{NOZ,NS,ENOO,MNSS} relating to possible quantum effects in brane world 
cosmology (BWC) models.
The ideas relate to the one brane Randall-Sundrum model \cite{RS2} and its 
generalisation to include a curved brane \cite{SMS,GT}.
 An interesting BWC scenario has been developed in \cite{KKS, HIME}, 
known as the bulk inflaton model, where it is possible to obtain
inflation on a single positive tension brane solely due
 to the effect of a  bulk gravitational scalar field. 
In this regard, the vacuum energy of the bulk scalar field
could have some affect on the cosmological evolution of the
brane, (depending on the size and sign of this quantity) and
is an interesting subject in its own right. 

Employing $\zeta$-function methods we evaluate the one-loop effective 
potential for a de Sitter brane embedded in a bulk 5-dimensional
anti-de Sitter spacetime. In fact, for this case, on the conformally related cylinder the 
effective potential reduces to the Casimir energy on $S^4$, see \cite{NOZ}. 
Thus, first we also evaluate the vacuum (Casimir) energy on the 
Euclidean Einstein universe $S^4\times R$. We focus on massless 
conformally coupled Majorana spinor fields and compare with the scalar field 
results. 
This is based on results from \cite{MNSS} which continues from
previous work \cite{NS}, for scalar fields. For references relating to flat brane 
calculations see those cited in \cite{NS}. 

In what follows we mention that there 
are no zero modes to deal with. This is because for fermion fields the relevant 
boundary conditions are mixed, meaning half the field components satisfy Neumann 
and the other half satisfy Dirichlet boundary conditions. This cancels any zero modes 
(by zero modes, we mean $n=0$ modes in our mode sum and not null eigenvectors).

\section{Casimir energy}

Before evaluating the vacuum energy, we must find the eigenvalues to be 
employed in the $\zeta$-function.
We start with the Euclidean metric suitable for de Sitter branes \cite{GS}
\begin{equation}
ds^2 =dr^2+\ell^2\sinh^2(r/\ell)d\Omega^2_4,
\label{orig}
\end{equation}
where $\ell=(-6/\Lambda_5)^{1/2}$ is the anti-de Sitter radius and
$d\Omega_4^2$ is the metric on the unit 4-sphere.
At the classical level, boundary or junction conditions relate the location of the 
single positive tension brane (located at $r_0$) to the brane 
tension $\sigma$ according to \cite{GS}
\begin{equation}
\sigma=\frac{3}{4\pi G_5 \ell}\coth(r_0/\ell).
\end{equation}
The quantum corrections will introduce other sources of stress-energy on the
brane which modify these relations. The metric can be written in conformal 
form
\begin{equation}
ds^2 =a^2(z)(dz^2+d\Omega^2_4)
\quad\qquad a(z)=\frac{\ell}{\sinh(z_0+|z|)}\,,
\label{metric}
\end{equation}
where the coordinates are chosen to have the positive tension brane at $z=0$.
The above form of the metric is conformal to a cylinder $I\times S^4$.
However, as in the bulk inflaton model, it may also be of interest to consider the 
compact space $S^1\times S^4$ \cite{GS}. If two branes are present then 
the other brane is placed at $|z|=L$, with the one brane limit given by 
$L\rightarrow \infty$.

For spin-$1/2$ fermion fields, $\psi$, the massless Dirac equation is 
automatically conformally covariant. We shall concentrate on Majorana spinors, 
see \cite{MNSS}. Taking the square of the Dirac operator to evaluate the effective 
action gives
\begin{equation}
W^M=-\frac14\log\det\Delta
\end{equation}
for Majorana fermions, where
\begin{equation}
\Delta=-\nabla^2+\frac14{}R^{(5)}
\end{equation}
The squared operator is not conformally invariant, but it is nevertheless 
still possible to relate the effective actions of fermions with conformally 
related metrics \cite{MOSS2}.

On the cylinder and noting that for conformal coupling all delta function contributions 
cancel,
\begin{equation}
\Delta=\left (-\partial_ z{}^2-\Delta^{(4)}_f+3\right),
\label{eigferm}
\end{equation}
where on $S^4$, $R^{(4)}=12$.
The eigenvalues of the spinor Laplacian, $\Delta^{(4)}_f$, on the 4-sphere 
are well known (e.g., see \cite{MOSS}) and are given by $(m+2)^2-3$. 
Half of the field components satisfy Dirichlet boundary conditions and half 
satisfy Neumann boundary conditions, commonly known as 
mixed boundary conditions. The eigenvalues of $\Delta$ are
\begin{equation}
\label{eig}
{\lambda_{n,m}^M}=\left(\frac{\pi n}{L}\right)^2+(m+2)^2\,.
\end{equation}
The degeneracy for the 8 component Majorana spinors is given by
\cite{MOSS}
\begin{equation}
d^M(m)=8\times\frac{1}{6}(m+1)(m+2)(m+3)
\label{degferm}
\end{equation}
where we take $n\in Z$. Then, the $\zeta$-function method can be employed 
to find the contribution to the effective action from the cylinder. 
The generalised zeta function is given by
\begin{equation}
\label{gfunc}
\zeta^M(s)=\sum_{m,n=0}^\infty d(m) \lambda_{m,n}^{-s}.
\end{equation}
with the one-loop effective action related to $\zeta(s)$ by (e.g., see 
\cite{MOSS, BD})
\begin{equation}
\label{effact}
W^M=\frac{1}{4}\,\zeta^{M\prime}(0)+\frac{1}{4} \zeta(0) \log \mu^2
\end{equation}
for Majorana fermions, where $\mu$ is the renormalisation scale. The 
effective potential is then obtained by dividing by the total volume.

As in \cite{NS} we first begin with the simpler case of the vacuum energy for the
Einstein universe $S^4\times R$. The evaluation of such quantities on 
$S^3\times R$  using an exponential cut off in the mode sum was performed by 
Ford \cite{FORD,FORD2}, for scalar and spinor fields respectively.
This is equivalent to finding the zero point energy of the equation
\begin{equation}
\partial_t^2\psi-\Delta_f^{(4)}\psi+\frac{1}{4}R^{(4)}\psi=0.
\end{equation}
The eigenvalues and degeneracy are given by Eq. (\ref{eig},\,\ref{degferm}), with 
$n=0$. From our definition of $W^M$ (Eq. (\ref{effact})\,) the vacuum energy for a 
Majorana field is defined as
\begin{equation}
E_0^M=-\frac 1 4 \zeta^M(-1/2)=
-\frac{1}{3}\left(\zeta(2s-3,2)-\zeta(2s-1,2)\right)_{s\rightarrow -1/2}=0,
\label{ein}
\end{equation}
where $\zeta(a,b)$ is the generalised (or Hurwitz) $\zeta$-function. 
This result is expected because in an odd number of dimensions there is no 
conformal anomaly \cite{BD}, i.e. as in the case of $S^2\times R$.

Now, we come to evaluate the effective potential for a one brane configuration on the
cylinder, where the discrete $n$ modes become continuous. 
It is then possible to show that
\begin{equation}
\zeta^M(s)=\frac{2L}{\pi}\int_0^\infty dk \sum_{m=0}^\infty d^M(m)
\left(k^2+(m+2)^2\right)^{-s},
\end{equation}
where the factor of $2$ is because we choose to undo the $Z_2$ symmetry as in 
\cite{GS}. For large $s$ we can interchange the order of the sum and the integral and 
perform the $k$ integration, 
\begin{eqnarray}
\zeta^M(s)&=&\frac{2L}{\pi}\,
\frac{\sqrt{\pi}}{2}\sum_{m=0}^\infty
\frac{\Gamma(s-1/2)}{\Gamma(s)}\;d^M(m)(m+2)^{1-2s},
\nonumber\\
&=&\frac{L}{\pi}\frac 4 3 \sqrt{\pi}\frac{\Gamma(s-1/2)}{\Gamma(s)}
\left(\zeta(2s-4,2)-\zeta(2s-2,2)\right),
\end{eqnarray}
where in the second step we have used simple algebra to rewrite the equation in 
terms of generalised (Hurwitz) $\zeta$-functions. The analytic continuation 
to $s=0$ is contained naturally in the definition of the $\zeta$-function, in this case.
For $s=0$ it is clear that $\zeta^M(0)=0$ because 
$\frac{1}{\Gamma(s)}=s+\gamma s^2+O(s^3)$,
where $\gamma$ is Euler's constant. Thus,
\begin{equation}
\label{1bravac}
\zeta^{M\prime}(0)=\frac{L}{\pi}\frac 4 3 \sqrt{\pi}\,\Gamma(-1/2)
\left(\zeta(-4,2)-\zeta(-2,2)\right)=0.
\end{equation}
Therefore, as for scalar fields \cite{NOZ,NS}, the one loop effective potential 
on the cylinder is zero in the one brane case. 
Just to see the relationship between the vacuum energy Eq. (\ref{ein}) and the 
effective potential, let us note that
\begin{equation}
W=-\frac 2 3 L\,\left(\zeta(-4,2)-\zeta(-2,2)\right)=0.
\end{equation}
Thus, apart from the extra factor of 2 due to our mode doubling, the effective 
potential agrees with the vacuum energy density on $S^4\times R$.

\section{Summary}

In this talk we have evaluated, on the conformally related cylinder, the effective potential for a 
de Sitter brane embedded in a bulk AdS$_5$ spacetime. We also evaluated the vacuum 
energy on the Euclidean Einstein universe $S^4\times R$ and found the two results 
to agree. However, there is a slight subtlety concerning our calculation, because 
when boundaries are present conformal transformations induce a 
non-trivial Jacobian in the path integral measure. This is taken care of by evaluating an 
extra term, known as the cocycle function \cite{MNSS}. Working on the cylinder, 
defined in Eq. (\ref{metric}), the cocycle function is found to vanish, because the 
half-cylinder volume is infinite, as first argued in \cite{NOZ}. However, using the actual 
metric, defined in Eq. (\ref{orig}), the cocycle function does not vanish because the 
volume of the Euclidean spacetime is finite, see \cite{MNSS}.

As the metric in Eq. (\ref{orig}) is the metric of physicality we conclude that the 
cocycle function must be included. However, there is still a close relationship between 
the vacuum energy density on $S^4\times R$ and the effective potential for a 
single $S^4$ brane in AdS$_5$.


Aside from the cocycle function, the fact that the vacuum energy on $S^4\times R$ 
with a mass or non-conformal curvature coupling will be non-zero implies that this 
will also be true of the de Sitter brane case. 
This considerably simplifies our analysis for the one brane case, which on the conformally 
related cylinder is equivalent to the evaluation of the Casimir energy on $S^4\times R$.
When applying the conformal transformation technique to the most general case, i.e. 
any given mass or curvature coupling, the potential is non-homogenous on the cylinder.
In this case, we must find the vacuum energy for a non-constant field 
configuration including background distributional sources.

\section{Acknowledgements}

W.N. acknowledges support from JSPS for Postdoctoral Fellowship for Foreign 
Researchers No. P01773.
W.S. is grateful to CONACYT of Mexico, Grant Number 116020, for financial 
support.
The work of M.S. is supported by Monbukagaku-sho Grant-in-Aid
for Scientific Research (S) No. 14102004.


\end{document}